\documentclass[fleqn,10pt]{olplainarticle}
\title{Sex Trouble: Common pitfalls in incorporating sex/gender in medical machine learning and how to avoid them}

\author[1]{Kendra Albert*}
\author[2]{Maggie Delano*}
\affil[1]{Harvard Law School, Cambridge, MA, USA}
\affil[2]{Swarthmore College, Swarthmore, PA, USA}


\begin{abstract}
False assumptions about sex and gender are deeply embedded in the medical system, including that they are binary, static, and concordant. Machine learning researchers must understand the nature of these assumptions in order to avoid perpetuating them. In this perspectives piece, we identify three common mistakes that researchers make when dealing with sex/gender data: ``sex confusion", the failure to identity what sex in a dataset does or doesn't mean; ``sex obsession", the belief that sex, specifically sex assigned at birth, is the relevant variable for most applications; and ``sex/gender slippage", the conflation of sex and gender even in contexts where only one or the other is known. We then discuss how these pitfalls show up in machine learning studies based on electronic health record data, which is commonly used for everything from retrospective analysis of patient outcomes to the development of algorithms to predict risk and administer care. Finally, we offer a series of recommendations about how machine learning researchers can produce both research and algorithms that more carefully engage with questions of sex/gender, better serving all patients, including transgender people.
\end{abstract}

\begin{document}

\flushbottom
\maketitle
\thispagestyle{plain}

\section{The Bigger Picture}
In recent years, there has been an increasing emphasis on the generation and sharing of datasets for machine learning research. When used for healthcare applications, these datasets often include demographic variables such as sex/gender, age, and race/ethnicity. However, these datasets often ignore the complexities of sex/gender and race/ethnicity, especially when using data from electronic health records. This can lead to worse health outcomes for anyone who does not fit the ``norm," especially transgender and non-white people. In this perspective piece, we review how assumptions around sex and gender have been embedded in medicine, and discuss how those assumptions influence machine learning research. We highlight both pitfalls and opportunities for machine learning researchers to better incorporate sex and gender data as appropriate. 

\section{Introduction}\label{introduction-k}

It is perhaps a clich\'e at this point to point out that like many industries, healthcare has become increasingly data-heavy and dependent. Many actors within health systems see great opportunity in both data gathered explicitly for research as well as the exhaust produced by everyday interactions between patients and doctors \citep{ghassemiReviewChallengesOpportunities2019}. Machine learning is one of the ways in which these data can be transformed - into algorithms that identify tumors \citep{chenMachineVisionassistedIdentification2022}, predict HIV risk \citep{marcusUseElectronicHealth2019}, or even warn when a patient in surgery might need an adjustment to their anesthesia \citep{lundbergExplainableMachinelearningPredictions2018}. 

As data becomes the scaffolding upon which new edifices are built, analyzing its own foundation is even more important. In this perspective, we look at one very particular aspect of medical data - the contingent and complicated relationship of variables like sex, gender, and sex/gender to the truth of bodies and identities. We explore how an insufficiently critical eye towards these complex variables can end up reifying the primacy of ``sex assigned at birth", a concept that has never done a particularly good job of capturing medically relevant information. We use examples of machine learning in healthcare, with a focus on electronic health records, to illustrate some common problems in analysis of sex/gender, and provide recommendations of how to use these variables in ways that support nuanced, meaningful findings that properly analyze sex/gender. 

Our message is a hopeful one - there are a variety of tools available to machine learning researchers and medical researchers to avoid the mistakes of the past, especially as more nuanced gender information is incorporated into standard electronic health records systems. But these tools require an understanding of what sex/gender are, and what they are not. 

\section{What's Wrong with Sex/Gender in Medicine?}\label{whats-wrong-with-sexgender-in-medicine}

Sex and gender are terms that refer to a variety of aspects of a person's physiology, how they see themselves, and how they interact with others and their environment. Sex typically refers to a person's gonads, anatomy, chromosomes, and hormone levels \citep{ainsworthSexRedefined2015,davisIntersexSocialConstruction2017}. Gender meanwhile typically refers to a person's gender identity (how they see themselves or experience their own gender), but also involves other factors such as how a person is perceived by others or experiences differential treatment related to their perceived gender \citep{shattuck2019sex,raparelliIdentificationInclusionGender2021,johnsonBetterScienceSex2007}. While it is often assumed that sex and gender are distinct, with sex being ``in the body'' and gender being ``in the mind,'' the two are actually deeply entwined \citep{springerCatalogueDifferencesTheoretical2012}. We will frequently use the term sex/gender to refer to the entwinement of these phenomena. Components of sex/gender and how a person interacts and is treated in a social world have deep impacts on their health and wellbeing \citep{reykFrontlineViewNational1990,tadiriMethodsProspectivelyIncorporating2021,knightGenderCardiovascularDisease2021,connellyImportanceGenderUnderstand2021}. Additionally, although a full treatment is outside the scope of this work, gender is also deeply entwined with race, with race mediating how people experience their own genders and how others gender them \citep{MyGenderBlack2017,jones2009shifting}.  

\begin{table}[tbp]
  \centering
  \caption{Definitions of sex, gender, and sex/gender related terms used in electronic health records. For other definitions of terms used in this paper, see Table \ref{tab:sex-gender-concepts}.}
    \begin{tabular}{p{7em}p{23em}p{7em}}
    \toprule
    \textbf{Term} & \multicolumn{1}{l}{\textbf{Definition}} & \textbf{Source} \\ \midrule
    Sex   & Sex refers to a person's status as male, female, or intersex based on biologic and physiologic characteristics. Sexes are usually assigned at birth based on simple visual inspection of the genitals of a newborn baby. & \cite{lauRapidReviewGender2020,goldmanTransPeopleBlood2020} \\
    Sex assigned at birth (SAAB) & Sex assigned at birth, usually based on simple examination of the genitals of a newborn baby. & \cite{lauRapidReviewGender2020,goldmanTransPeopleBlood2020} \\
    Legal Sex & Sex as defined by legal documents, such as birth certificate, passport, driver's license, or health care card. Many transgender people may have different legal sexes across contexts. & \cite{lauRapidReviewGender2020} \\
    Gender & Gender can refer to personal and/or interpersonal components, including gender identity (man, woman, non-binary, agender, genderfluid), and also gender roles and stereotypes. Gender is often assumed to be binary, static, and concordant with sex (i.e. a baby assigned male at birth is and will always be a man, a baby assigned female at birth is and will always be a woman). & Present work \\
    Administrative gender & A term used in most existing electronic health record systems that refers to either one's sex or gender recorded and used for administrative purposes, such as billing. Also known in as legal gender in some EHR systems. & \cite{lauRapidReviewGender2020} \\
    Sex/gender & A term used in place of sex to emphasize the entanglement of sex and gender. & \cite{springerCatalogueDifferencesTheoretical2012} \\
     \bottomrule
    \end{tabular}%
  \label{tab:sex-gender-terms}%
\end{table}%

\subsection{Sex/gender as binary, static,
concordant}\label{sexgender-as-binary-static-concordant}

Historically, sex/gender has been assumed to be binary, static, and concordant \citep{ainsworthSexRedefined2015,davisIntersexSocialConstruction2017,fausto-sterlingSexingBodyGender2000}. Binary means that there were two options for a person's sex (male and female), and two options for a person's gender (man or woman). Once assigned male or female at birth, typically by examining a newborn's external genitalia, a person is presumed to be male or female for their entire life (static). Anyone assigned male at birth is also presumed to be a man, and anyone assigned female at birth is presumed to be a woman (concordant). 

In practice, sex and gender are not binary, static over time, or necessarily perfectly aligned. There are people who have chromosomal, gonadal, anatomical, and/or secondary sex characteristics that do not align with the defaults for male or female sex. Some people with such characteristics may use the term intersex for themselves, but some do not. Estimates suggest intersex people make up between 1.7 and 4 percent of the population (though an exact number is difficult to quantify, see \cite{jonesIntersexStudiesSystematic2018}). Similarly, there are people with genders other than man or woman (e.g. non-binary, genderqueer, genderfluid), and some people have no gender (agender). Some people's gender changes over short or long periods of time. A person's sex and gender also need not be aligned; transgender people are assigned a sex at birth that does not align with their gender identity. 

Despite these complexities, the false assumptions of binary, static, and concordant sex/gender are deeply ingrained in our medical system \citep{tannenbaumSexGenderAnalysis2019a,richardsonSexContextualism2022,schiebingerIntegratingSexGender2021}. This negatively impacts not just those with minority sex/gender, but also fails to serve anyone who doesn't exactly fit the average (which is almost everyone in some way or another).

Trouble with sex/gender has historically manifested in clinical care and research in three main ways: 1) ignoring complexities entirely and treating everyone the same, 2) excluding people from minority groups or comparing minority groups to a ``default'' (cisgender white men), or 3) presuming and essentializing differences that lead to stereotypes and discrimination against minority groups \citep{garcia-sifuentesReportingMisreportingSex2021,richardsonSexContextualism2022}. In recent years, funding agencies such as the National Institutes of Health (NIH) in the United States have attempted to address scenarios 1) and 2) by requiring the inclusion of women and racial minorities in clinical trials \citep{NOTOD02001AMENDMENTNIH,NOTOD18014RevisionNIH} and analyzing sex and/or gender as a dependent variable in biological studies \citep{NOTOD15102ConsiderationSex,governmentofcanadaHowCIHRSupporting2018}. However, with their emphasis specifically on ``women,'' ``female health'' and ``sex differences'', these interventions actually reinforce a binary, static, and concordant framing of sex/gender, often excluding intersex, transgender, and non-binary people. A lack of consideration of differences in sex/gender leads to lack of knowledge about the experiences of and appropriate treatment for sex/gender minorities, and incorrect, denied, or delayed treatment \citep{knightGenderCardiovascularDisease2021,stroumsaPowerLimitsClassification2019,spadeResistingMedicineRemodeling2003,todayTransgenderAdultReference2019,rosendaleAcuteClinicalCare2018,goldsteinWhenGenderIdentity2017, maysTransfusionSupportTransgender2018}.

\subsection{Problems with Sex as Ground Truth: Sex Obsession, Sex Confusion, and Sex/Gender Slippage} \label{sex-confusion-problems-with-sex-as-ground-truth}
In recent years, there has been a laudable effort to improve the experience of LGBTQIA+ patients through the addition of gender-related fields to EHR systems \citep{lauRapidReviewGender2020,keuroghlianElectronicHealthRecords2021,thompsonImplementationGenderIdentity2021}. While the introduction of gender identity and other fields in EHR such as pronouns and preferred names can help facilitate a more welcoming environment for trans, non-binary, and gender non-conforming people, the addition of these fields has been mostly limited to patient-facing care. In many cases, ``sex" is still assumed to be the ``ground truth," both by doctors and by researchers.  We call this assumption ``sex obsession"—the idea that fundamentally, sex, usually sex assigned at birth, corresponds to some particular fact about the bodies of people with that designation, and is thus the most relevant factor to analyze when attempting to produce generalizable knowledge.

Sex obsession often manifests as studies that are purportedly about gender heavily leaning on sex assigned at birth. For example, Pelletire et al. created a composite gender score in order to evaluate its association with risk factors in patients with premature acute coronary syndrome \citep{pelletierCompositeMeasureGender2015}. However, they used sex as a dependent variable when creating their gender index, which assumes everyone in the cohort is cisgender and endosex (i.e. not intersex). As the results from the authors themselves show, there is a wide variety of ``gender" scores for ``male" and ``female" participants. Additionally, characterizing gender on a one-dimensional scale from masculine to feminine ignores the complexities of gendered experience.

Sex obsession leads to what we call ``sex confusion"—a failure to actually develop an understanding of what sex means in a particular dataset or context. The meanings of sex and gender in any previously gathered dataset are context-dependent and will change over time. For example, EHR systems have begun to distinguish between gender identity (how a person experiences their own gender and identifies), administrative gender (the sex/gender used for administrative purposes such as billing), and sex assigned at birth (the sex assigned based on examination of a newborn's genitalia)\citep{lauRapidReviewGender2020, deutschCollectionGenderIdentity2014}. However, when only a ``sex" field is available, as has been historically the case in EHR data, that field could actually correspond to any one of these three things. Prior to the introduction of gender identity fields within medical records, many transgender people may have changed their demographic sex information in order to increase their chances of being treated respectfully by medical doctors. Some may still be forced to do so in order to get access to particular forms of care or legal tools \citep{currahWeWonKnow2008}. As an example of how this plays out, that means that the demographics of what a particular sex marker at a particular hospital means may end up depending on what the local rules are for name changes, and be subject to shifts as those rules are modified. The advent of ``gender identity" as a separate field in EHRs may minimize some of the pressure to change administrative gender, but this does not mean that the ``sex" field is now clean and easy to interpret, as sex assigned at birth has always been in itself socially constructed and contingent \citep{davisIntersexSocialConstruction2017,fausto-sterlingSexingBodyGender2000}. 

Sex obsession and sex confusion can lead to situations where false assumptions are made based on a patient's sex marker, particularly that all people with those designations have a particular body part, secondary sex characteristic, or hormone level \citep{alpertRethinkingSexassignedatbirthQuestions2021}. It is very common for clinicians and researchers to assume that an ``M" in the demographics section means not just a penis, but testosterone levels in a particular range, higher muscle mass, or a particular pelvic build. These assumptions do not hold up particularly well in practice, and not just for transgender and gender non-conforming people. As an example, post-menopausal cisgender women have decreased levels of estrogen as compared to pre-menopausal cisgender women \citep{MenopauseTransitionEndocrine2005}. Although both groups may show up with an ``F" as their sex marker in a medical record, and indeed, have had genital erectile tissue under a certain size at birth, the assumption individuals belongings to both groups would have similar expected estrogen levels would be incorrect. 

When using sex as a demographic variable in a research study, researchers sometimes assume that sex in EHR systems not only corresponds to a set of chromosomes, anatomy, hormone levels, etc. but also to gender, leading to sex/gender slippage. This is made worse by the fact that datastores themselves confuse sex characteristics and gender in their database definitions, as Kronk et al. point out \cite{kronkTransgenderDataCollection2021}. It is thus unsurprising that researchers lack clarity on whether they are talking about sex or gender. For example, in Feller et al., the authors switch back and forth between describing the information retrieved from the EHR system as gender and sex \citep{fellerUsingClinicalNotes2018}. Such slippage can also occur in analysis. For example, in Ancochea et al., researchers gathered sex data from EHRs, and then, when attempting to explain their results, discuss disproportionate childcare obligations on women \citep{ancocheaEvidenceGenderDifferences2021}. This dynamic is typical across the literature we reviewed. Sex may be used because of presumptions around its accuracy (or frankly, the fact that the field is almost always filled out) but researchers postulate gendered explanations for their results. It is not wrong to turn to an analysis of gender for biomedical phenomena, as gender is lived in the body, having impacts on health. But ``sex" demographic data in medical records may not, in many cases, correspond to a patient's experience of gender. 

\begin{table}[tbp]
  \centering
  \caption{Definitions of sex, gender, and sex/gender related terms used to characterize the use of sex/gender in medical machine learning and possible improvements. For other definitions of terms used in this paper, see Table \ref{tab:sex-gender-terms}.}
    \begin{tabular}{p{7em}p{23em}p{7em}}
    \toprule
    \textbf{Term} & \multicolumn{1}{l}{\textbf{Definition}} & \textbf{Source} \\ \midrule
Sex/gender slippage & The free substitution of sex for gender and gender for sex, and the presumed equivalence of sex and gender. & \cite{albertThisWholeThing2021a} \\
    Sex essentialism & An approach to research that ``perpetuate(s) a view of sex differences as principally biological and not also cultural, leading to a potentially dangerously inaccurate understanding of the causes of sex disparities." & \cite{richardsonSexContextualism2022,epsteinInclusionPoliticsDifference2009} \\
    Sex contextualism & An approach to research that ``recognizes the pluralism and context-specificity of operationalizations of 'sex' across experimental laboratory research" & \cite{richardsonSexContextualism2022} \\
    Sex confusion & A characterization of the uncertainty of the accuracy of the ``sex" field in an EHR system, which could represent SAAB, legal sex, administrative gender and/or gender identity. & Present work \\
    Sex obsession & The prioritization of the binary M/F sex variable over the gender variable or other sex/gender related variables in clinical and machine learning research. & Present work \\
    Organ/Anatomic Inventory & An accurate record of what organs a patient may or may not have & \cite{lauRapidReviewGender2020,rosendaleAcuteClinicalCare2018, deutschElectronicMedicalRecords2013} \\
    phenotyping & The identification of particular patients who might have certain ``characteristics of interest." & \cite{bandaAdvancesElectronicPhenotyping2018} \\
    data richness & Richness moves beyond the binary presence or absence of a condition to timing, degree, severity, cause, and relationship to factors like behavior, etc. & \cite{hripcsakHighfidelityPhenotypingRichness2018} \\      \bottomrule
    \end{tabular}%
  \label{tab:sex-gender-concepts}%
\end{table}%

\subsection{Sex/Gender in Machine Learning Research}\label{issues-present-in-machine-learning-studies}
The issues laid out with sex variables above have profound implications for all data analysis done at scale in medicine, including analysis that uses machine learning. Machine learning researchers often pick up from where healthcare and society have ``left off'', encoding these biases and then introducing their own. This means that ML cannot easily divorce itself from the assumptions made in this previous stage, especially when relying on EHR or existing datasets. An individual researcher might be able to make careful decisions about how to understand sex/gender within the context of their work, knowing which characteristics are key to the questions they are asking. EHR or other standardized data, on the other hand, is one-size-fits-no-one. Researchers using EHR data cannot go back and ask each clinician how the ``sex" field was populated, and due to variable adoption of more inclusive EHR practices, may not even be aware of when or how sex and gender identity were considered separately. But because sex variables are almost universally filled out within EHR systems, they can seem appealing for ML purposes \citep{khoElectronicMedicalRecords2011}. Notably, none of the papers using machine learning on EHR data that we could find discussed how sex data was gathered or what it meant, and only a handful even identified that sex was a binary variable.

\subsubsection{Diving Deeper into EHR Data Concerns}

EHR data has been used by machine learning researchers in a variety of contexts, including for retrospective analysis to better understand health phenomena \citep{wanyanContrastiveLearningImproves2021}, to identify patients for potential treatments or interventions \citep{krakowerAutomatedIdentificationPotential2016}, and/or to guide clinical care in real time \citep{lundbergExplainableMachinelearningPredictions2018}. In this section, we discuss common pitfalls around sex/gender when performing machine learning on EHR data.

Sex/gender slippage and sex confusion were common across papers we studied. For example, Walsh et al. use EHR data to analyze suicide risk, incorporating ``gender" data from EHRs because ``demographics such as age and gender are known risk factors for suicidal behavior" \citep{walshPredictingRiskSuicide2017a}. In the study, ``gender" was found to be a significant factor in prediction of suicide attempts. Given that that paper was published in 2017 and Vanderbilt Medical system did not start collecting specific gender identity data until that year \citep{dingModelImprovingHealth2020}, it seems difficult to imagine that gender information was actually used. This paper, like many, demonstrates a sex/gender slippage of the type described in our previous work \cite{albertThisWholeThing2021a}. Such a slippage here is especially notable because gender identity and gendered experiences likely do matter for suicidality, as transgender and gender-nonconforming people have significantly higher risk of suicide. Slippages between sex, gender, and sex/gender, and a failure to understand the underlying actual data may lead to inaccurate conclusions about suicide attempts as a gendered phenomenon, and may specifically fail to recognize patterns around transgender suicidality.

Strikingly, studies that particularly focus on gender can exhibit similar issues. As discussed briefly above, Ancochea et al. used the unstructured free text of EHRs from a hospital system in Spain to look at differences in management and treatment for COVID-19 \citep{ancocheaEvidenceGenderDifferences2021}. The research seems to use sex information from the demographic field of the EHR system to determine sex/gender. (We say seems to because the exact method they used for determining sex/gender is not clear from the methodology.) However, the results and discussion theorizes a number of different potential explanations for differences between male and female patients. Some, as discussed above, vary based on gender (e.g. women being more likely to have primary care giving responsibilities), and some vary based on things like ovarian hormones. A more robust analysis of what the ``sex" variable means in this context could allow for a more detailed analysis of how these variables interact with the population in question. 

Studies like these can serve as building blocks, potentially resulting in a failure to produce effective knowledge translating to other contexts. Within medicine, machine learning has long been used for ``phenotyping" - that is, the identification of particular patients who might have certain ``characteristics of interest" \citep{bandaAdvancesElectronicPhenotyping2018}. Such algorithms can be used for recruitment of patients for clinical trials, cohort or case control studies on already existing data, or even clinical decision support. Although phenotyping is similar in some ways to exploratory analyses of the type discussed above, the excitement about using phenotyping algorithms in shaping subsequent care means that potential failures to account for sex/gender complications could result in people being left out of both research \emph{and} potentially helpful medical care. Although most phenotyping that uses machine learning currently only uses sex/gender to balance populations, phenotypes specific to demographics are likely on the horizon, as such phenotypes have already been developed using non-ML methods \citep{chenBuildingBridgesElectronic2015,estiriEvolvingPhenotypesNonhospitalized2021}. 

The slippage and confusion are most concerning in circumstances where machine learning is used to directly guide clinical care. For example, Lundberg et al. use explainable ML models to predict the prevention of hypoxaemia during surgery, allowing for anesthesiologists to adjust administration of anesthesia accordingly \citep{lundbergExplainableMachinelearningPredictions2018}. In general, explainable models that have been thoroughly vetted and reviewed by working clinicians, like the one produced by Lundberg et al., reduce some of the risks that can be involved in black box models where sex/gender may end up playing a significant role as a feature without the knowledge of researchers, doctors, or patients. However, even these very well-documented models use sex/gender from EHRs as an input, without fully analyzing what that variable might mean. And, to the extent that such models might also use real-time test or device data to guide interventions, sex/gender may matter because some tests use sex/gender offsets or sex/gender-based reference intervals \citep{goldsteinWhenGenderIdentity2017}. If the model is built such that the EHR ``sex" field is the relevant one for determining such thresholds/offsets, it could be wrong for whole subsections of the population without any clinician being the wiser. 

To our knowledge, none of the key features in Lundberg rely on backend sex/gender offsets. However, it is only the interpretability of their algorithm that allows us to  analyze it. To the extent that machine learning systems may at some point be used to produce so called ``dimensionless parameters" \citep{ganBispectralIndexMonitoring1997,albertThisWholeThing2021a}, the failure to specify how questions of sex/gender might affect these numbers could have deleterious effects on transgender patients. The failure to allow for data richness can be offset somewhat in an individual patient relationship, where a doctor may be able to look beyond whatever the demographics say. But ML-developed algorithms are the opposite of that. In short, the closer that machine learning based on EHR data gets to prescribing clinical care without allowing for a care team to determine whether the assumptions made as part of algorithmic development are correct for the patient, the more significant the risks are to individual patients who may ``deviate" from the expectations of the people who built the systems.

\subsubsection{Case Study: HIV and PrEP}
The risks of ``sex obsession" are especially significant in settings where sex, gender and sex/gender are known to play a significant role in outcomes, such as when transgender and gender non-conforming people are overrepresented in the populations studied. For example, a number of studies have used machine learning (or electronic risk scoring) to determine when a patient should be offered pre-exposure prophylaxis (PrEP), medicine that can reduce the risk of HIV infection see \citep{krakowerAutomatedIdentificationPotential2016,marcusUseElectronicHealth2019,fellerUsingClinicalNotes2018,ridgwayWhichPatientsEmergency2018}). HIV is often transmitted through sexual activity, and it's hard to imagine an area of human behavior where questions of sex/gender play a larger role. It might be tempting to reduce risk variance in HIV transmission to either sex assigned at birth or genitals, but such a framing is inaccurate. A 2017 discussion paper from Here for Health in Australia highlights a number of potential differences in how transgender people might experience risk related to HIV transmission compared with cisgender people. Potential differences include the differential proprieties of cis women's vaginas and transfeminine people's neo-vaginas and potential effects of testosterone on genital lubrication for transmasculine people with vaginas \citep{reykFrontlineViewNational1990}. And of course, sex practices themselves are highly gendered.

Many studies that attempt to do predictive modeling in this space, which can be used to determine when it might be appropriate to do a follow up call for HIV risk or PrEP counseling, rely on EHR sex data. For example, Ridgway et al. say in their abstract that their risk model incorporates gender, but the EHR data is ``sex" \citep{ridgwayWhichPatientsEmergency2018}. \footnote{This distinction is especially interesting because a table of people who received an electronic alert based on the output of the algorithm they developed includes not just ``male" and ``female" as sexes, but also a lone ``transgender female."}  Likewise, ML studies relied on ``sex" data \citep{fellerUsingClinicalNotes2018, marcusUseElectronicHealth2019}. Marcus et al. also used EHR sex as an input variable, with male sex being one of the six features that they defined as most predictive. The algorithm produced could not identify cases amongst ``females" - presumably meaning people with the EHR sex ``F." How that corresponds to any particular set of body parts or experiences is unclear, as the paper goes on to discuss how transgender women might be best characterized by a prediction model for females. Of course, transgender women are women, but not all transgender women may have their sex marked as F within an EHR system. Marcus et al. also note that they could not ``assess sensitivity by gender identity" because there were too few incident diagnoses among transgender patients \citep{marcusUseElectronicHealth2019}. The most charitable read is that this means that there were too few diagnoses for people who had the ``gender identity" field filled out in their EHR, but the language also seems to unfortunately imply cisgender people do not have gender identities.

Notably, there are better models. Krakower et al., the earliest work that suggests using ML to predict HIV risk and guide PrEP distribution, does not use a binary sex variable, likely because the data used was from Fenway Health, an organization specializing in LGBTQ healthcare. However, it is unclear which features are most predictive of HIV risk of the 168 variables extracted from the EHR, so it is difficult to draw any conclusions about how sex/gender was accounted for in the results \citep{krakowerAutomatedIdentificationPotential2016}. 

\subsubsection{Other Issues Related to Sex/Gender}
In this analysis of sex confusion within machine learning, we focus on the ways in which EHR ``sex" fields can generate false assumptions and how researchers focus on ``sex" to the exclusion of other relevant variables. We observe that sex confusion is more common than sex obsession, in part because there is such a lack of clarity on what researchers think the variable they are studying is. But of course, there are other issues present. Transgender and gender non-conforming people, due to stigma, the casual transphobia of the medical field, and lack of access to resources that might result in being unable to seek care in a country with uneven access to medicine, are likely underrepresented in any data set that relies on historical medical data \citep{dingModelImprovingHealth2020, hinesTheyJustDon2019}. Additionally, many studies rely on natural language processing tools to handle unstructured data present within notes. This creates additional difficulties, as the notes field within EHRs often contains the musings of healthcare professionals on questions of sex/gender. They also may use keywords that assume particular relationships of sex/gender and behavior, for example, the use of ``homosexual" in the context of HIV risk. Such issues are harder to identify but vital to eliminate if ML-based knowledge production becomes more normalized.

None of this analysis focuses on the way that particular models or studies might over or underrepresent sex/gender as a causal variable. Frankly, at this point, such analysis is not possible because the current failures are too significant. The historical failure of EHR systems to account for sex/gender in nuanced ways, the sex/gender slippage in medical literature, and the ways in which the base assumption that the complexity of sex/gender differences can be eliminated by looking at sex all make drawing conclusions about bias within models nigh on impossible.

\section{Recommendations}\label{recommendations-m}

While assumptions about sex/gender are deeply embedded in the medical system, we believe that there are concrete steps machine learning researchers can take to better incorporate sex/gender in their work. In this section, we provide recommendations related to increasing "data richness" \citep{hripcsakHighfidelityPhenotypingRichness2018} and on avoiding common methodological pitfalls.

\subsection{Focus on data richness}\label{focus-on-data-richness}

Our most important recommendation is to, whenever possible, not treat sex/gender as a stable, coherent category. Focus instead of increasing the ``richness'' of representations
\citep{hripcsakHighfidelityPhenotypingRichness2018}. This includes:

\begin{itemize}
    \item Differentiating between sex assigned at birth and current sex in the EHR system
    \item Considering the presence/absence of particular organs using an organ/anatomic inventory \cite{lauRapidReviewGender2020,rosendaleAcuteClinicalCare2018, deutschElectronicMedicalRecords2013}
    \item Considering variables besides binary sex/gender, including hormone status, gender identity, etc.
\end{itemize}

In order to focus on data richness, it is important to triage an analysis based on the clinical implications of the work, and work with clinicians, data experts and patients. For studies with a lot of different variables, triage your emphasis based on how the models will be used. We divide this into three main categories:

\begin{enumerate}
\item purely retrospective study not used for clinical decision making
\item used in routine clinical decision making, but adjusted as needed by
  clinician
\item ML or other algorithm acts as final decision maker without intervention from clinicians
\end{enumerate}

Scenarios with higher category numbers have the most direct clinical impact and should be evaluated the most rigorously. Researchers should test their models for differential performance whenever possible. It is worth noting that unlike in many ML implementations outside of healthcare \citep{kourouMachineLearningApplications2015}, it does not appear that medical researchers are suggesting using live data produced by machine learning in the context of direct clinical care. Given the current state of machine learning knowledge and risk management, this is fortunate. However, to the extent that this changes, careful handling of sex/gender data is vital. This is especially true for algorithms that do not produce interpretable results or worse, result in dimensionless parameters that can create special risks for populations that do not conform to the assumptions of their creators. Our findings serve to reinforce the need for interpretability and clarity around feature significance in machine learning in the medical context.

\subsection{Avoid Methodological
Pitfalls}\label{avoid-methodological-pitfalls}

As we've discussed throughout this paper, sex/gender data need to be interpreted carefully. When starting a research study, look to previous research (if it exists) about the possible roles of sex/gender \citep{tannenbaumSexGenderAnalysis2019a,lacasseConductingGenderbasedAnalysis2020}. Do not assume sex/gender is relevant, but similarly, do not assume that it is not \citep{richardsonSexContextualism2022}. This can be a challenge, but ``threading the needle'' is important to better understanding the role of sex/gender in any particular study. We've observed four major methodological pitfalls to avoid: 1) not working with data experts, 2) not describing how sex/gender data were collected, 3) incorporating gender in a way that does not consider the complexities of sex/gender, and 4) misrepresenting sex/gender differences.

\subsubsection{Work With Data Experts}

Researchers should always work with clinicians and data experts to avoid the pitfalls around data bias or data being used in unintuitive ways. It is especially important to find a clinician who has worked with these data specifically, as it might not always be clear how something is actually used in clinical practice. EHR data can be used for multiple or unintuitive purposes, such as the preferred name field, which is sometimes also used to store information like pronouns or phone numbers. Also consider working directly with patients when feasible.

\subsubsection{Clearly Describe How Sex/Gender Data Were Collected and Used}

One of the biggest challenges in evaluating research papers and their use of sex/gender is that in the vast majority of cases papers do not even disclose how sex/gender data were collected. It is assumed to be so obvious and clear as to not be stated explicitly. However, as we've discussed above, how sex/gender data are collected has impacts on their accuracy. When conducting a research study, state clearly where the sex/gender variables used in your study came from, and how they were collected. If the variables available shifted over the course of the time that data was gathered, as if a gender identity field was added to an EHR system,  include details about when and how the change occurred and how your analysis accounts for these changes. For example, if using the ``sex" field from a patient's medical record, discuss who populated that field and how it is being used in the research. If there is information you do not know, state this clearly as as a limitation of the study and discuss potential impacts on the results.

\subsubsection{Incorporate Gender Carefully}\label{incorporate-gender-carefully}

Researchers of gender and health typically break gender down into four dimensions, including gender identity, relations, roles, and institutionalized gender \citep{johnsonBetterScienceSex2007,tadiriMethodsProspectivelyIncorporating2021}. Different components and variables related to gender will be relevant depending on the research context. It is also important not to conflate sex and gender, or assume that sex is a ``ground truth" where as gender is in people's heads. This oversimplifies the relationship between sex and gender and erases transgender, non-binary, intersex, and gender non-conforming people. 

Steps to incorporate gender into an analysis include: 1) identifying potentially relevant gender-related variables, 2) defining measured outcomes, and 3) selecting gender-related variables from an existing EHR or dataset or developing a protocol to collect new data. When incorporating gender, consider its multi-dimensional nature and which variables might be relevant for the research context. This can be challenging as there may be many different variables to consider. One way to simplify the analysis is to reduce multiple gender-related variables into a composite score by summing or averaging \citep{tadiriMethodsProspectivelyIncorporating2021}. However, given the number of variables, it can be difficult to determine how to weigh them relative to one another. Methods such as exploratory factor analysis or principle component analysis could be used to reduce the number of variables and/or construct the appropriate composite score. If conducting a statistical analysis using a gender-related variable, consider also how gender might impact a measured outcome. For example, gender could act as a main effect, directly impacting an outcome, or it could act as an interaction term (e.g., with sex). One challenge might be the lack of availability of gender related-variables in the EHR or data set. One approach to mitigate this would be to incorporate other available indexes. However, these indices may have assumptions about sex/gender that diminish their utility, such as excluding the experiences of transgender, non-binary, and intersex people.

\subsubsection{Conduct Statistical Tests
Appropriately}\label{conduct-statistical-tests-appropriately}

When considering the possible influence of sex/gender in research, it is important to conduct the appropriate statistical tests to ensure sex/gender differences are not misreported. A recent study from Garcia-Sifuentes and Maney examined the reporting of sex differences in over 100 articles in the biological sciences and found that while sex differences are increasingly reported, statistical evidence for those claims were missing or based on incorrect statistical methods \citep{garcia-sifuentesReportingMisreportingSex2021}.  There were two pitfalls the authors identified: failing to test for an interaction effect, and pooling without looking for differences. When looking at the effect of a ``treatment" or intervention, an effect in one sex but not another was frequently reported as a ``sex difference." However, this does not imply the presence of a sex difference. Instead, one must test for the interaction between sex and the treatment. The other pitfall was pooling based on sex/gender without testing first whether there are statistically and clinically significant differences between groups, or pooling even when differences had been clearly established. Similar issues could be possible in some machine learning contexts, for example those that involve evaluating the impact of some sort of treatment or intervention \citep{ridgwayWhichPatientsEmergency2018}, or deciding whether to include sex/gender related variables as parameters when training a model. Our review of the machine learning literature has found that a binary measure of sex/gender is often included without subsequent evaluation of its influence on the results. 

\section{Conclusion}

Machine learning plays an increasingly important role in modern healthcare. Machine learning researchers need to learn about how important demographic variables such as sex and gender are used in medical data sets, especially electronic health records (EHR). In this paper, we provide an overview of historical assumptions about sex and gender in medicine. We highlight that sex and gender have been viewed as binary, static, and concordant, with sex and gender frequently used interchangeably (sex/gender slippage). Ignoring the complexities of sex, gender and the entwinement of them (i.e. sex/gender) can lead to significant ``sex confusion" when attempting to use these data.  Researchers must be careful not to continue to reify sex-assigned-at-birth as ``ground truth" over other variables (sex obsession). The failure to think about what sex is and does makes it difficult to apply algorithms in any case where an individual person might deviate from dimorphic norms, whether that person is transgender, intersex, has undergone surgery or treatment to remove sex organs, or just doesn't quite conform to the normal curves (no pun intended). With the advent of gender identity being added to EHRs, data richness and literature that describes how to avoid common methodological pitfalls when it comes to sex/gender provide promising tools for machine learning researchers to move forward.

Many of the problems we have described are not unique to the phenomenon of sex/gender \citep{paulladaDataItsDis2021}. However, because of the historical beliefs about sex and what we have termed sex obsession, even researchers who are relatively savvy and thoughtful about the origins of their data and its (dis)contents, to borrow the phrasing of Paullada et al., can make the mistake of thinking that ``sex" is one of the things in a dataset that doesn't need further explanation or unpacking. Make no mistake. Despite its position in the demographics section, sex is more like ``family history" than it is like ``age." The same can, to some extent, be said for race and ethnicity \citep{lettConceptualizingContextualizingOperationalizing2022}. 

The nature of modern machine learning systems and the lack of interpretability to clinicians can make the dynamics described above even more concerning. As of yet, we have not seen machine learning systems deployed in medical settings that continuously train on patient data. But the machine learning and even basic algorithmic scoring systems that have been produced, especially those in areas like PrEP counseling or referrals, take ``sex" at face value, resulting in less accurate and less than helpful results. It is vital that as machine learning researchers look to the low-hanging fruit of EHRs to make ``meaningful use" of the data within, they pay full attention to the deep contextual knowledge required to interpret what might seem like the most straight-forward of variables.

\section{Acknowledgements}\label{acknowledgements}
The authors would like to thank David Arney, Heather Lane, and Jordan Nestor.

\bibliography{Smacks}

\section{Biographies}

Maggie Delano is an Assistant Professor at Swarthmore College, where she teaches computer engineering and inclusive engineering design. Her research focuses on the development of a wearable monitoring system for patients with Congestive Heart Failure. She also writes about and researches better design, including the use of sex/gender in medicine and human subjects testing in machine learning. More details about Maggie can be found on her website: https://www.maggiedelano.com/

Kendra Albert is a public interest technology lawyer whose research interests include gender, machine learning, and power. They serve as a clinical instructor at the Cyberlaw Clinic at Harvard Law School, where they teach students to practice law by working with pro bono clients. Kendra also has taught classes on transgender rights in the Program on Studies of Women, Gender, and Sexuality at Harvard College and Harvard Law School. They serve on the board of the ACLU of Massachusetts and the Tor Project, and provide support as a legal advisor for Hacking // Hustling. More at https://kendraalbert.com/.

\end{document}